\documentstyle[12pt]{article}
\newcommand{\drawsquare}[2]{\hbox{%
\rule{#2pt}{#1pt}\hskip-#2pt
\rule{#1pt}{#2pt}\hskip-#1pt
\rule[#1pt]{#1pt}{#2pt}}\rule[#1pt]{#2pt}{#2pt}\hskip-#2pt
\rule{#2pt}{#1pt}}

\newcommand{\Yfund}{\raisebox{-.5pt}{\drawsquare{6.5}{0.4}}}
\newcommand{\Ysymm}{\raisebox{-.5pt}{\drawsquare{6.5}{0.4}}\hskip-0.4pt%
        \raisebox{-.5pt}{\drawsquare{6.5}{0.4}}}
\newcommand{\Yasymm}{\raisebox{-3.5pt}{\drawsquare{6.5}{0.4}}\hskip-6.9pt%
        \raisebox{3pt}{\drawsquare{6.5}{0.4}}}
\newcommand{\Ythree}{\raisebox{-3.5pt}{\drawsquare{6.5}{0.4}}\hskip-6.9pt%
        \raisebox{3pt}{\drawsquare{6.5}{0.4}}\hskip-6.9pt
        \raisebox{9.5pt}{\drawsquare{6.5}{0.4}}}
\newcommand{\Ysthree}{\raisebox{-.5pt}{\drawsquare{6.5}{0.4}}\hskip-0.4pt%
        \raisebox{-.5pt}{\drawsquare{6.5}{0.4}}\hskip-0.4pt
        \raisebox{-.5pt}{\drawsquare{6.5}{0.4}}}

\newcommand{\beq}{\begin{equation}}
\newcommand{\eeq}{\end{equation}}
\newcommand{\beqa}{\begin{eqnarray}}
\newcommand{\eeqa}{\end{eqnarray}}

\def\vbr{\vphantom{\sqrt{F_e^i}}}

\newcommand{\Group}[2]{{\hbox{{\sl #1}($#2$)}}}
\newcommand{\U}[1]{\Group{U\kern0.05em}{#1}}
\newcommand{\SU}[1]{\Group{SU\kern0.1em}{#1}}
\newcommand{\SL}[1]{\Group{SL\kern0.05em}{#1}}
\newcommand{\Sp}[1]{\Group{Sp\kern0.05em}{#1}}
\newcommand{\SO}[1]{\Group{SO\kern0.1em}{#1}}

\newcommand{\jref}[4]{{\it #1} {\bf #2}  (#4) #3}
\newcommand{\NPB}[3]{\jref{Nucl.\ Phys.}{B#1}{#2}{#3}}

\newcommand{\PRD}[3]{\jref{Phys.\ Rev.}{D#1}{#2}{#3}}

\begin{document}

\pagestyle{empty}

\begin{titlepage}
\def\thepage {}        

\title{`t Hooft Anomaly Matching for QCD}

\author{John Terning\thanks{e-mail: terning@alvin.lbl.gov}\\
\\
\baselineskip=10pt
{\small \it Department of Physics,
University of California,
Berkeley, CA 94720}\\
{\small \it  and}\\
{\small \it Theory Group, Lawrence Berkeley National Laboratory, Berkeley, CA
94720}}

\baselineskip=12pt

\date{}

\maketitle
\vspace*{-90mm}
\noindent
\makebox[11cm][l]{}UCB-PTH-97-32\\
\makebox[11cm][l]{}LBNL-41477
\vspace*{100mm}

\abstract{
I present a set of  theories which display non-trivial `t Hooft anomaly
matching for QCD with $F$ flavors.  The matching theories are non-Abelian
gauge theories with ``dual" quarks and baryons, rather than the purely
confining theories of baryons that `t Hooft originally searched for.  The
matching gauge groups are required to have an $F\pm 6$ dimensional
representation. Such a correspondence is reminiscent of Seiberg's duality
for supersymmetric (SUSY) QCD, and these theories are candidates for
non-SUSY duality. However anomaly matching by itself is not sufficiently
restrictive, and duality for QCD cannot be established at present. At the
very least, the existence of multiple anomaly matching solutions should
provide a note of caution regarding  conjectured  non-SUSY dualities.}

\vfill
\end{titlepage}

\baselineskip=18pt
\pagestyle{plain}
\setcounter{page}{1}

\section{Introduction}
Many years ago, `t Hooft proposed searching for confining gauge theories with
composite massless fermions \cite{tHooft}. The primary tool in such a search is
the `t Hooft anomaly matching condition: the requirement that the global
anomalies of a proposed low-energy effective theory equal those
of the original ultraviolet theory.
The requirement that no chiral symmetries are broken makes
such a matching highly non-trivial.  The further requirement that any number of
fermion flavors can be decoupled (i.e. that adding  mass terms and integrating
out flavors can be accounted for in the low-energy effective description by
integrating out all composites containing that flavor) puts another stringent
constraint on proposed solutions to the  `t Hooft problem.   In his pioneering
work \cite{tHooft}, `t Hooft showed that for a vector-like \SU{3} gauge theory
with fundamental quarks, a solution could only be found  for the case of two
flavors. He also showed that there were no solutions for \SU{5} theories with
fundamental quarks.  More general searches were performed \cite{Albright} and a
handful of possible solutions were found for chiral and vector-like theories.

More recently Seiberg \cite{Seib} has revolutionized our understanding of
supersymmetric (SUSY) gauge theories.  In addition to finding confining SUSY
gauge theories with massless composite fermions (and their SUSY partners),
Seiberg also found theories with dual descriptions in terms of a different
gauge group with different matter content. These dual theories can have
trivial or non-trivial infrared fixed points. Seiberg's work obviously raises
the question  of whether such dual gauge descriptions persist in non-SUSY
theories. Various authors \cite{softbreak} have considered the effects of
adding soft-SUSY breaking mass
terms and progress has been made when such masses are much smaller than the
intrinsic scale
of the gauge theory, however little is known about what happens when the SUSY
breaking masses are increased to be larger than the intrinsic scale. Recently
D-brane constructions \cite{dbrane} have also led to speculations about
non-SUSY dualities.  The main evidence for the conjectured duality is anomaly
matching, but, as I will argue below, that by itself is insufficient to
demonstrate duality.

Under what circumstances would it be reasonable for another gauge theory to
provide an alternate description of the infrared physics of QCD? Certainly when
the number of quark flavors, $F$,  is sufficiently large so as to produce an
infrared fixed point there is no a priori objection to such a duplicate
description since such a conformal gauge theory has no particle interpretation.
These fixed point theories cannot be said to possess a low-energy effective
theory in the conventional sense, but there may be other conformal gauge
theories that describe the same fixed point. Banks and Zaks \cite{Banks} have
shown that by analytically continuing in $F$ it is possible to establish such
an infrared fixed point for QCD in perturbation theory for $F$ below and
sufficiently close to $33/2$.  Presumably such fixed point behavior persists as
$F$ is reduced down to some critical value $F^{\rm crit}$.  Assuming that
chiral symmetry breaking\footnote{Or equivalently \cite {rainbow} that the
anomalous dimension of  the quark mass operator equals one.} marks the end of
the fixed
point regime, Appelquist, Wijewardhana, and I estimated \cite{QCDFP} this
critical value to be $F^{\rm crit} \approx 4 N_c = 12$.  Lattice Monte Carlo
studies \cite{lattice} suggest that $F^{\rm crit} > 8$. Whatever the precise
value of  $F^{\rm crit}$ is, for $F$ in the range $F^{\rm crit} < F < 33/2$ it
seems worthwhile to consider a generalized `t Hooft problem: is there a gauge
theory coupled to composite massless fermions that matches the anomalies of
QCD?  Such a matching theory would be a candidate for a dual description of
QCD.

\section{An Anomaly Matching Theory}

The theory I wish to study is QCD with $F$ flavors: there is an \SU{3} gauge
group with $F$ left handed quarks
$Q_L$ and $F$ right-handed quarks $Q_R$. This theory has the anomaly-free
global symmetry $\SU{F}_L \times \SU{F}_R  \times \U{1}_B$ i.e. a chiral flavor
symmetry and vector-like baryon number.
The fermion content (with global charges) is given in Table 1.

\begin{table}[htbp]
\centering
\begin{tabular}{|c||c|ccc|}\hline
field & \SU{3} & $\SU{F}_L$  & $\SU{F}_R$ & $\U{1}_B$
\\\hline \hline
$Q_L$ & \Yfund & \Yfund & {\bf 1}  & ${{1}\over{3}}\vbr$   \\
$Q_R$ &  \Yfund & {\bf 1}  & \Yfund & ${{1}\over{3}}\vbr$  \\
\hline
\end{tabular}
\label{QCD}
\parbox{4in}{\caption{Fermion content of QCD with $F$ flavors.}}
\end{table}

A solution of the generalized `t Hooft problem requires a gauge theory which
matches all the global anomalies of QCD with $F$ flavors.
The fermion content of a model that accomplishes this is displayed in
Table 2. The matching theory contains some left-handed and right-handed ``dual"
quarks which belong to an $F-6$ dimensional representation of the gauge group
$G$.  There are also gauge singlet, flavor antisymmetric fermions
$A_{L,R}$ and $B_{L,R}$ that have the
correct quantum numbers to correspond to baryonic composites of the original
quarks.  Note that the baryons labeled $B$ are the large flavor (chiral)
analogs of  the
proton and neutron, while the $A$ baryons are the analogs of the
$\Lambda$ baryon.

\begin{table}[htbp]
\label{Match}
\centering
\begin{tabular}{|c||c|ccc|}\hline
field & $G(F-6)$ & $\SU{F}_L$  & $\SU{F}_R$ & $\U{1}_B$
\\
\hline\hline
$q_L$ & \Yfund &  \Yfund & {\bf 1} &    ${{F-2}\over{F-6}} \vbr$ \\
$q_R$ &\Yfund & {\bf 1} & \Yfund &${{F-2}\over{F-6}} \vbr$ \\
$A_L$ & {\bf 1} & \Ythree & {\bf 1}  & $1\vbr$  \\
$A_R$ & {\bf 1} &  {\bf 1} &  \Ythree & $1 \vbr$\\
$B_L$ & {\bf 1} & \Yfund & \Yasymm &  $1 \vbr$   \\
$B_R$ & {\bf 1} &  \Yasymm &\Yfund &$1 \vbr$ \\
\hline
\end{tabular}
\parbox{4in}{\caption{Fermion content of the matching theory.}}
\end{table}

The anomaly matching equation for the $\SU{F}_L^3$ and $\SU{F}_R^3$ anomalies
is:
\beq
3 = F-6 + {{1}\over{2}}(F-3)(F-6) +  {{1}\over{2}}F(F-1) - F(F-4)~,
\eeq
while the equation for the
$\SU{F}_L^2 \U{1}_B$ and  $\SU{F}_R^2 \U{1}_B$  anomalies is:
\beq
1 = (F-6) {{F-2}\over{F-6}}  + {{1}\over{2}}(F-2)(F-3) +  {{1}\over{2}}F(F-1) -
 F(F-2)~.
\eeq
Note that the matching of  the $\U{1}_B^3$ and mixed gravitational anomalies is
trivial since both theories are vector-like with respect to  $\U{1}_B$, while
the matching of the $\SU{F}_R^2 \U{1}_B$ amounts to a definition of the
$\U{1}_B$ charge of the dual quark.

The requirement of anomaly matching by itself does not determine whether the
gauge group $G(F-6)$ is
\SU{F-6}, \SO{F-6}, or (for even $F$) \Sp{F-6}.  In fact all that is determined
is that this group have an $F-6$ dimensional representation, so there are even
more possibilities.  Certainly the matching theory is not sensible for $F <7$.
For $F=7$, the gauge representation is one dimensional, so the only possibility
is a $U(1)$ gauge group.  Since such a $U(1)$ theory is  free in the infrared,
such a description would not fulfill the theoretical prejudice mentioned above
that  duals are reasonable for the case of non-trivial infrared fixed points
(also recall that lattice studies
\cite{lattice} suggest $F^{\rm crit} > 8$.).   If the gauge group $G(F-6)$ is
in fact \SU{F-6} then the matching theory is asymptotically free for $F \ge 8$,
while if $G(F-6)$ is really \SO{F-6} then it is asymptotically free for $F \ge
10$.  Thus if the matching theory is to provide a dual description of QCD, this
gives a weak preference to the gauge group \SO{F-6}.

\section{Decoupling a Flavor}
Since the anomaly matching is independent of $F$ the decoupling of a flavor is
straightforward.  It is instructive however to consider what type of dynamics
could produce the
correct decoupling in the matching theory.  A thorough understand of this
decoupling would be required in order to establish that the matching theory is
actually a dual description of the same physics. Thus I will consider in
some detail one
possible mechanism for the case that $G(F-6)$ is identified with \SU{F-6}.

Adding a mass term to the QCD Lagrangian for the $F$-th flavor,
\beq
{\cal L}=m \bar{Q}_{R}^F Q_L^F + \rm{h.c.}~,
\eeq
decouples one flavor, so the number of flavors $F$ is reduced by one in the
infrared.  In the matching theory
the gauge group must be broken to a group with a representation with dimension
$(F-7)$ and mass terms are needed for fermions carrying an index $F$.
Such mass terms can be achieved with the introduction of the scalar fields
displayed in Table 3.

\begin{table}[htbp]
\label{Scalars}
\centering
\begin{tabular}{|c||c|ccc|}\hline
field & $G(F-6)$ & $\SU{F}_L$  & $\SU{F}_R$ & $\U{1}_B$
\\
\hline\hline
$\phi$ & \Yfund &   \Yfund & \Yfund  &    ${{-4}\over{F-6}} \vbr$ \\
$M$ & {\bf 1} & $\overline{\Yfund}$ & \Yfund  & $0 \vbr$  \\
\hline
\end{tabular}
\parbox{4in}{\caption{Possible scalar content for the matching theory.}}
\end{table}
If the matching theory has the following Yukawa interactions:
\beq
{\cal L}= \lambda_1 M \bar{q}_R q_L
+\lambda_2 M (\bar{B}_{R} A_L  + \bar{A}_R B_{L}  ) + \lambda_3 M \bar{B}_{L}
B_{R}+\lambda_4 \phi (\bar{B}_{R} q_L + \bar{B}_{L} q_R ) + \rm{h.c.} ~,
\eeq
then the  correct mass terms are generated when $M^{FF}$ and $\phi^{FF}_{F-6}$
have non-zero vevs.  Note that if the gauge group of the matching theory is
\SU{F-6}, then the vev of $\phi$ engenders the correct Higgs
mechanism to break the  gauge group to \SU{F-7}. It is also
interesting to note that the gauge singlet field $M$ has the correct quantum
numbers to correspond to a mesonic bound state of the original quarks. Of
course, in a non-SUSY theory there is no reason for scalars to be light in the
absence of fine-tuning.  If the scalars are required for a putative dual
description of QCD, perhaps masses tuned to be of order $\Lambda_{QCD}$ are
sufficient for decoupling purposes rather than the more stringent requirement
of masslessness.

\section{A Second Anomaly Matching Theory}

The fermion content of a second model that provides a solution of the
generalized `t~Hooft problem is displayed in Table 4. The matching theory
contains some left-handed and right-handed ``dual" quarks which belong to an
$F+6$ dimensional representation of the gauge group $G$.  There are also gauge
singlet, flavor symmetric fermions $S_{L,R}$ and $T_{L,R}$ which have the
correct quantum numbers
to correspond to baryonic composites of the original quarks.

\begin{table}[htbp]
\label{Match2}
\centering
\begin{tabular}{|c||c|ccc|}\hline
field & $G(F+6)$ & $\SU{F}_L$  & $\SU{F}_R$ & $\U{1}_B$
\\
\hline\hline
$q_L$ & \Yfund &  $\overline{\Yfund}$ & {\bf 1} &    $-{{F+2}\over{F+6}} \vbr$
\\
$q_R$ &\Yfund & {\bf 1} & $\overline{\Yfund}$ &$-{{F+2}\over{F+6}} \vbr$ \\
$S_L$ & {\bf 1} & \Ysthree & {\bf 1}  & $1\vbr$  \\
$S_R$ & {\bf 1} &  {\bf 1} &  \Ysthree & $1 \vbr$\\
$T_L$ & {\bf 1} & \Yfund & \Ysymm &  $1 \vbr$   \\
$T_R$ & {\bf 1} &  \Ysymm &\Yfund &$1 \vbr$ \\
\hline
\end{tabular}
\parbox{4in}{\caption{Fermion content of the matching theory.}}
\end{table}

The anomaly matching equation for the $\SU{F}_L^3$ and $\SU{F}_R^3$ anomalies
is:
\beq
3 = -(F+6) + {{1}\over{2}}(F+3)(F+6) +  {{1}\over{2}}F(F+1) - F(F+4)~,
\eeq
while the equation for the
$\SU{F}_L^2 \U{1}_B$ and  $\SU{F}_R^2 \U{1}_B$  anomalies is:
\beq
1 = -(F+6) {{F+2}\over{F+6}}  + {{1}\over{2}}(F+2)(F+3) +  {{1}\over{2}}F(F+1)
-  F(F+2)~.
\eeq

In contrast to the previous matching model, the gauge group $G(F+6)$ is
asymptotically free for any value of $F$, for any one of the three possible
identifications:  \SU{F+6},  \SO{F+6}, or \Sp{F+6}. The baryons labeled $T$ are
additional chiral,
large flavor analogs of  the proton and neutron, while the $S$ baryons are the
analogs of the
(orbitally excited) spin-${{1}\over{2}}$  decuplet baryons.

The two matching theories I have presented here are the only solutions to the
generalized `t Hooft problem (the search for a gauge
theory coupled to composite massless fermions that matches the anomalies of
QCD) subject to two additional simplicity conditions: that each
distinct composite baryon appears at most once, and that the dimension of the
gauge representation for the
``dual"  quark be linear in $F$ with coefficient $1$. This statement can be
proved
simply by enumerating all the possible baryonic operators\footnote{I have also
assumed
that a composite made completely of left-handed quarks is itself left-handed,
for it to be a  right-handed state would require a larger ``orbital angular
momentum".}.
These simplicity
criterion may seem somewhat arbitrary, but they are motivated by the form of
Seiberg's SUSY dualities. Determining whether they are actually necessary
conditions will require a deeper understanding of duality. I have also looked
for
anomaly matchings for larger gauge groups like \SU{5}, but have not found
any simple generalizations of the models presented here \cite{future}.
Loosening these criteria allows further solutions for QCD and other gauge
theories.  For example dropping the condition that the coefficient of $F$ be
equal to $1$,  matching solutions can be constructed for an \SU{5} gauge theory
with $F$ flavors involving both flavor symmetric and antisymmetric baryons and
``dual"  quarks in gauge representations of  dimension $19F \pm 5$. Dropping
the constraint of linearity in $F$ completely allows a number of matching
solutions
for QCD where dimension of the gauge representation for the  ``dual"  quark is
quadratic in $F$.
\section{Conclusions}
I have presented two solutions to the generalized `t Hooft problem for QCD.
These are the only solutions subject to two constraints of simplicity.
It is clear that I have not established that the matching theories are dual in
the sense of Seiberg.  In the absence of more powerful consistency checks, it
is impossible to tell whether both, either, or neither of the matching
solutions gives a correct description of QCD physics for the range of flavors
$F^{\rm crit} < F < 33/2$ (i.e. the infrared fixed point phase).    Thus, while
the anomaly matching between QCD and the models I have described is
intriguing, it may only be a mathematical curiosity rather than a consequence
of a duality.  More generally, one should be skeptical about any conjectured
non-SUSY duality that relies only on anomaly matching for support.

What are the future prospects for establishing non-SUSY dualities? Further
progress
will almost certainly rely upon establishing new consistency checks. In the
absence of SUSY,  anomaly matching gives no information about the boson content
of the theory. In SUSY theories there is generally a finite dimensional moduli
space of
inequivalent vacua that must match between dual pairs, however for non-SUSY
theories there is a unique vacuum (up to global symmetry transformations), so
the correspondence is trivial. Furthermore, in Seiberg's analysis he was able
to require that
the dual description provided a correct description of the physics for any
number of flavors.  Thus a powerful consistency check was provided by
integrating out one flavor at time and seeing that the mapping of the quark
mass term to the dual description lead to the correct physics in each case. For
non-SUSY theories however, phase transitions are to be expected as $F$ is
varied (since  there are no constraints from holomorphy), so even
if a dual description were established in the infrared fixed point phase it may
not be possible to reduce $F$ and produce the correct dual description (i.e.
the
chiral Lagrangian) for the chiral symmetry breaking phase.  In the absence of
new analytic consistency checks, the question of duality for QCD will probably
require lattice Monte Carlo calculations for a definitive answer.  Even with
lattice calculations the analysis will not be straightforward.
In the case of theories with non-trivial infrared fixed points, a candidate
dual
theory cannot be considered as a low-energy effective theory but rather as
merely an equivalent description of the low-energy physics.  Thus a test of
duality for theories with non-trivial infrared fixed points will require two
lattice computations, one for each of the dual pair, and detailed comparison of
the infrared physics observed in each.

\section*{Acknowledgements}
I would like to thank  Nima Arkani-Hamed, Nick Evans, Ann Nelson, Martin
Schmaltz, and Adam Schwimmer for helpful discussions.
This work was supported the National Science Foundation under grant
PHY-95-14797 and also partially supported by
the Department of Energy under contract DE-AC03-76SF00098.



\begin{thebibliography}{99}
\bibitem{tHooft}G. `t Hooft, {\em Recent Developments in Gauge Theories}
(Plenum Press, 1980) 135; reprinted in {\em Unity of Forces in the Universe
Vol. II}, A. Zee ed. (World Scientific 1982) 1004.

\bibitem{Albright}C.H. Albright, {\em Phys. Rev.} {\bf D24} (1981) 1969, and
references therein.

\bibitem{Seib}
N. Seiberg, \PRD{49}{6857}{1994}, hep-th/9402044;
 \NPB{435}{129}{1995}, hep-th/9411149;
for a review see K. Intriligator and N. Seiberg,
RU-95-48, IASSNS-HEP-95/70, hep-th/9509066.

\bibitem{softbreak}O. Aharony, et. al. {\em Phys. Rev.} {\bf D52} (1995) 6157,
hep-th/9507013;
N. Evans, et. al. {\em Nucl. Phys.} {\bf B456} (1995) 205, hep-th/9508002.

\bibitem{dbrane}A. Brandhuber et. al.,
hep-th/9704044.

\bibitem{Banks}T. Banks and A. Zaks, {\em Nucl. Phys.} {\bf B196} (1982) 189.

\bibitem{QCDFP} T. Appelquist, J. Terning, and L.C.R. Wijewardhana,  {\em
Phys. Rev. Lett.} {\bf 77} (1996) 1214, hep-ph/9602385.

\bibitem{rainbow}H. Georgi and A. Cohen, {\em Nucl. Phys.} {\bf B314} (1989) 7.

\bibitem{lattice}J. B. Kogut and D. K. Sinclair, {\em Nucl. Phys.} {\bf B295}
[FS21] (1988) 465, F. Brown et. al., {\em Phys. Rev.} {\bf D46} (1992) 5655.

\bibitem{future}J. Terning, work in progress.

\end{thebibliography}
\end{document}